\documentclass[twocolumn]{autart} 
\usepackage{url}
\usepackage{mathptmx,amsmath,amssymb}

\newcommand{\ud}{\mathop{}\!\mathrm{d}}
\newcommand{\cond}{\,|\,}
\newcommand{\Half}{\textstyle \frac{1}{2}}
\newcommand{\To}{\text{:}}
\newcommand{\Transp}{\mathsf{T}}
\newcommand{\hP}{\hat{P} }
\newcommand{\hQ}{\hat{Q} }
\newcommand{\hG}{\hat{G} }

\DeclareMathOperator{\Normal}{N}
\DeclareMathOperator{\Tr}{tr}
\DeclareMathOperator{\Diag}{diag}

\begin{document}

\begin{frontmatter}
\runtitle{Automatic numerical differentiation}
\title{Automatic numerical differentiation
by maximum likelihood  estimation
of state-space model\thanksref{footnoteinfo}}
 \thanks[footnoteinfo]{This paper was not presented at any IFAC 
meeting. Corresponding author R. Pich\'e. Tel.\ +358 40 8490174.}

\author[TUT]{Robert Pich\'e}\ead{robert.piche@tut.fi}
\address[TUT]{Tampere University of Technology, Tampere, Finland}
\begin{abstract}
A linear Gaussian state-space smoothing algorithm is presented for
 estimation of derivatives from a sequence of noisy measurements.
The  algorithm uses numerically stable square-root formulas,
 can handle simultaneous independent measurements
and non-equally spaced abscissas, 
and can compute state estimates at points between the data abscissas.
The state space model's parameters, including driving noise intensity, 
measurement variance,
and initial state, are determined from the given data sequence
 using maximum likelihood estimation
 computed using a expectation maximisation iteration.
In tests with synthetic biomechanics data,
the  algorithm has equivalent or better accuracy
compared to other
automatic numerical differentiation algorithms.
\end{abstract}

\end{frontmatter}

\section{Introduction}

Numerical differentiation (ND) of a sequence of noisy measurements
 is an important  problem in data analysis.
For example, one may want to estimate velocity and acceleration from 
a sequence of displacement measurements.
The problem has been well studied;
comparative surveys of ND algorithms 
 include~\cite{pezzack77,brown92,damico92,corradini93,giakas97,walker98,ahnert07,puglisi15}.

Because differentiation  amplifies noise,
catastrophically so when the sampling rate is high,
an effective ND method must trade off data fidelity
with noise smoothing.
In most ND algorithms, the trade-off is governed by 
one or more user-defined parameters, 
variously called regularisation, smoothing, or bandwidth (cutoff frequency) parameters.
Some ND algorithms are ``automatic'', 
in the sense that they
determine the smoothing parameters 
for a given time series 
without knowledge of the true signal values.
The surveys \cite{corradini93,giakas97}
assess several automatic ND  algorithms.

Numerical differentiation can be approached
as a standard state space estimation problem 
with continuous-time dynamics and discrete-time measurements.
 In the Kalman fixed-lag smoother of  Fioretti and Jetto~\cite{fioretti89,fioretti94},
the state space dynamic model is a
 multiply-integrated stationary Wiener process,
 and
the measurement error is an
 additive stationary discrete-time Gaussian white noise.
In the target tracking literature this family of state-space models is known as
the polynomial motion model~\cite[\S6.2]{barshalom01},
of which the constant velocity model is the best known example.


The ND algorithm  presented here is also based on the  state space model
of the multiply-integrated stationary Wiener process.
Derivatives are estimated using
fixed-lag Rauch-Tung-Striebel smoothing 
implemented with numerically stable square-root formulas.
The algorithm can treat independent simultaneous measurements 
and non-equally-spaced abscissas,
and supports evaluation at abscissas other than data points (``dense output'').
A maximum likelihood (ML) estimate
of all the state space models' parameters,
namely the initial state,
driving noise intensity, and measurement noise variance,
is computed  using an extension of the expectation-maximisation (EM) algorithm
for  state space model identification~\cite{shumway82,gibson05}.
A \textsc{Matlab} implementation of the algorithm is freely available for download\footnote{
\url{https://se.mathworks.com/matlabcentral/fileexchange/xxxxxxx}}.



\section{Algorithm}

\subsection{Signal model}

The underlying signal is assumed to be 
the $(d-1)$-fold integral of a Wiener process.
The linear stochastic differential equation is
\[
    \ud x = F x\,\ud t +  q L \ud w 
\] 
where $w$ is the standard Wiener process,
 the underlying signal 
 is the first component of the $d$-component state vector $x$,
its first derivative 
is the second component, etc.,
and
\[ F= \begin{bmatrix} 
    0 & 1 & 0 & 0 & \cdots & 0\\ 
    0 & 0 & 1 & 0 & \cdots & 0\\  
    \vdots & &&&& \vdots \\
        0 & 0 & 0 &  0 & \cdots & 1 \\
    0 & 0 & 0 &  0 & \cdots & 0 \end{bmatrix}
    , \quad 
    L=\begin{bmatrix} 0 \\ 0\\  \vdots \\ 0 \\ 1 \end{bmatrix} .
\]
The  parameter $q>0$ is the intensity (spectral density) of the driving white noise.

The abscissas for the discrete-time
state space model are denoted $t_k$ for $k=1,2,\ldots$;
the  sequence $t_1,t_2,\ldots$ is assumed to be monotonically increasing.
Denoting  $x_k=x(t_k)$  and  $\Delta_{k}=t_{k+1}-t_{k}$,
the discrete-time dynamic model is
a linear state space model driven by additive discrete Gaussian white noise
\begin{equation}   \label{ssmProcess}
   x_{k+1} |x_{k}  \sim \Normal(A_{k}x_{k},Q_{k}) , \quad k=1,2,\ldots,
\end{equation}
where $N(\,\cdot\,,\,\cdot\,)$ denotes a Gaussian distribution with given mean and covariance,
the dynamic transition matrix is 
\[
 A_{k}=\exp(F\Delta_k)=
 \begin{bmatrix}  1 & \Delta_{k}  & \frac{1}{2!} \Delta_k^2 & \cdots & \frac{1}{(d-1)!} \Delta_k^{d-1}  ¤ \\
  0 & 1 & \Delta_k & \cdots & \frac{1}{(d-2)!} \Delta_k^{d-2} \\ 
  \vdots & &&& \vdots \\
  0 & 0  & 0 &  \cdots  & 1\end{bmatrix},
\]
and the driving noise covariance is 
$
    Q_k =q  \bar{Q}_k$,
where 
\begin{align*}
 & \bar{Q}_k =
    \int_0^{\Delta_k} \exp(F(\Delta_k-\tau))LL^\Transp  \exp(F^\Transp (\Delta_k-\tau)) \ud \tau
    \\ &=
   \Diag(\begin{bmatrix} \Delta_k^{\frac{2d-1}{2}} \\ \vdots \\ \Delta_k^{1/2} \end{bmatrix})
    \begin{bmatrix}    
    \frac{1}{2d-1} & \cdots & \frac{1}{4}  & \frac{1}{3}  \\[1ex]
      \frac{1}{2d-2}  & \cdots & \frac{1}{3}   & \frac{1}{2}  \\[1ex]
      \vdots & & & \vdots \\
        \frac{1}{d}  & \cdots & \frac{1}{2}  &  1
    \end{bmatrix}
     \Diag(\begin{bmatrix} \Delta_k^{\frac{2d-1}{2}} \\ \vdots \\ \Delta_k^{1/2} \end{bmatrix})
.
\end{align*}

At each abscissa, there are $n_k$ scalar measurements, denoted 
$y_{k,1},\ldots,y_{k,n_k}$.
Each measurement is modelled as the signal value plus 
independent additive zero-mean Gaussian  noise,
that is,
\begin{equation}   \label{ssmMeasurement}
    y_{k,j}|x_{k} \sim \Normal(H x_k,R) , \quad j=1,\ldots,n_k,
\end{equation}
where $H=[1,0,\cdots,0]$ and $R$ is the variance.


\subsection{Fixed-Interval Smoothing   \label{sec:smoothing}}

Let $x_{k|j}$ denote the state conditioned on the measurements  at times $t_1,\ldots,t_j$.
For the linear Gaussian state space model \eqref{ssmProcess} and \eqref{ssmMeasurement},
and a Gaussian prior distribution 
\begin{equation}   \label{ssm_initial}
    x_{1|0} \sim \Normal(m_{1|0},P_{1|0}),
\end{equation}   
all posterior states $x_{k|j}$ are jointly Gaussian.
\emph{Fixed-interval smoothing} is
the computation of
the mean and covariance of the states $x_{1|T},\ldots x_{T|T}$,
given the model parameters
\[  \theta= [q, R, m_{1|0}, P_{1|0}] \]
and  the measurements $y_{1:T}=[y_{1,1},\ldots,y_{T,n_T}]$.
The Rauch-Tung-Striebel (RTS) smoother computes these states sequentially,
with a forward pass (a Kalman filter) that processes the measurements,
followed by a backward pass.
For better numerical stability,
 the QR factorisation-based square root RTS algorithm of  \cite{gibson05}
 is used, as follows.

The \emph{forward pass}  consists of  two stages that are carried out for each $k=1,2,\ldots,T$.
Before the beginning of the forward pass,  $P_{1|0}^{1/2}$, 
the lower triangular Cholesky factor of $P_{1|0}$, is computed.
The first stage, the \emph{measurement update},
is the computation of the parameters of the filtering distribution
$
   x_{k|k} \sim  
   \Normal(m_{k|k}, P_{k|k}^{1/2}P_{k|k}^{\Transp /2})
$
 by the formulas
\begin{align*}
   & m_{k,0|k} = m_{k|k-1}  ,\;  P^{1/2}_{k,0|k} = P_{k|k-1}^{1/2}   \\
   & \text{for }j=1,\ldots,n_k\text{ do}   \\
   & \quad     \mathcal{R}  = \text{triangular factor of  QR decomposition of }
   \\ & \mbox{}\qquad \qquad
     \begin{bmatrix}  
         R^{1/2} & H P_{k,j-1|k}^{1/2} \\ 0 & P_{k,j-1|k}^{1/2} 
      \end{bmatrix}^\Transp   \\  
        & \quad   S_{k,j}  = \mathcal{R}_{1,1}^\Transp  \mathcal{R}_{1,1}\\ 
   & \quad K_{k,j}  =  \mathcal{R}_{1,2:d+1}^\Transp   \mathcal{R}_{1,1}^{-\Transp} \\
   & \quad P_{k,j|k}^{1/2}  = \mathcal{R}_{2:d+1,2:d+1}^\Transp  \\
   & \quad v_{k,j} = y_{k,j}-Hm_{k,j-1|k}   \\
   & \quad m_{k,j|k} =  m_{k,j-1|k}+K_{k,j} v_{k,j}  \\
   & \text{end do}  \\
  & m_{k|k}= m_{k,n_k|k},\;
     P_{k|k}^{1/2} = P_{k,n_k|k}^{1/2}    
\end{align*}

The second stage, the \emph{dynamic update}, 
is the computation of the parameters of the one-step prediction distribution
$
   x_{k+1|k} \sim 
    \Normal(m_{k+1|k}, P_{k+1|k}^{1/2}P_{k+1|k}^{\Transp /2})
$.
The formulas for the dynamic update  are
\begin{align*}
m_{k+1|k}&=A_k m_{k|k}   \\
 \mathcal{R} & = \text{triangular factor of  QR decomposition of }
    \\ & \mbox{}\qquad \qquad
     \begin{bmatrix}  
     P_{k|k}^{\Transp/2} A_k^\Transp \\ Q_k^{1/2} 
      \end{bmatrix}^\Transp 
     \\  
      \quad P_{k+1|k}^{1/2}  &= \mathcal{R}_{1\To d,1\To d}^\Transp 
\end{align*}
This stage is omitted for $k=T$.

In the \emph{backward pass},
the parameters of the joint smoothing distribution
\[ 
\begin{bmatrix} x_{k+1|T} \\ x_{k|T} \end{bmatrix}
\sim \Normal\Bigl(   
\begin{bmatrix} m_{k+1|T} \\ m_{k|T} \end{bmatrix},
\begin{bmatrix} P_{k+1|T} & P_{k+1|T}G_k^\Transp  \\ G_kP_{k+1|T} &P_{k|T} \end{bmatrix}
\Bigr)
\] 
 are computed sequentially for $k=T-1,\ldots,1$;
the  smoothing distribution is then $x_{k|T}\sim \Normal(m_{k|T},P_{k|T})$.
The backward pass formulas are
\begin{align*}
P_{k|k} &= P_{k|k}^{1/2}P_{k|k}^{\Transp/2} \\
P_{k+1|k}^{-1} &= P_{k+1|k}^{-\Transp/2} P_{k+1|k}^{-1/2} \\
G_k &=  P_{k|k} A_k^\Transp  P_{k+1|k}^{-1}    \\
 m_{k|T} &= m_{k|k}+G_k(m_{k+1|T}-m_{k+1|k}) \\
 \mathcal{R} & = \text{triangular factor of  QR decomposition of }
    \\ & \mbox{}\qquad \qquad
     \begin{bmatrix}  
P_{k|k}^{\Transp/2}A_k^\Transp & P_{k|k}^{\Transp/2} \\
            Q_k^{\Transp/2} &       0 \\
            0    &        P_{k+1|T}^{\Transp/2}   G_k^\Transp
            \end{bmatrix} \\
P_{k|T}^{1/2} &=   \mathcal{R}_{d+1\To 2d,d+1\To 2d}
\end{align*}

\subsection{Estimation of model parameters \label{sec:estimation}}

The maximum likelihood estimate of
the  model parameters  
$ \theta= [q, R, m_{1|0}, P_{1|0}]$
is the maximiser of the likelihood $p(y_{1:T} | \theta)$, 
or equivalently
 the minimiser of the ML cost function
\[
  \phi(\theta) = - \log p(y_{1:T} | \theta).
\]
For fixed $\theta$, the cost function can be computed inside the Kalman filter
(the first stage of the forward pass of the smoothing algorithm)
 using 
\begin{equation}   \label{eq:phi}
   \phi(\theta)={ \textstyle  \frac{1}{2}}\sum_{k=1}^T \sum_{j=1}^{n_k}
    \bigl( \log\det (2\pi S_{k,j}) + v_{k,j}^\Transp  S_{k,j}^{-1}v_{k,j} \bigr).
\end{equation}

In the Expectation-Maximisation (EM) method 
the ML estimate is found by iteratively maximizing a lower bound on the likelihood.
An EM method for state-space model parameters 
that uses a smoother to marginalise the state variables
is presented in~\cite{shumway82,gibson05}.
This method needs to be extended for
the ND state space model,
which has varying dynamic model matrices
and a single-parameter process noise matrix;
this is done in the appendix.
The EM parameter update formulas are
\begin{subequations} \label{EMformula}
\begin{align}
& q= \frac{1}{(T-1)d} \sum\limits_{k=1}^{T-1} \Tr (\hQ_k \bar{Q}_k^{-1} ),
\;
R =\frac{1}{N} \sum\limits_{k=1}^T  \sum\limits_{j=1}^{n_k}  \hat{R}_{k,j}
,\\
& m_{1|0} =\hat{m}_{1|T}
, \;
P_{1|0} =\hP _{1|T} ,
\end{align}
\end{subequations}
where 
\begin{subequations}   \label{QhatRhat}
\begin{align}   
   \hQ_k &= 
       [I,-A_k] \Bigl(  \begin{bmatrix} \hat{m}_{k+1|T}\\ \hat{m}_{k|T} \end{bmatrix} 
         \begin{bmatrix} \hat{m}_{k+1|T}\\ \hat{m}_{k|T} \end{bmatrix} ^\Transp ,
      \nonumber   \\
      & \qquad \mbox{}+  \begin{bmatrix} \hP _{k+1|T} &  \hP _{k+1|T}\hG_k^\Transp  \\ 
      \hG_k  \hP _{k+1|T} &  \hP _{k|T} \end{bmatrix} 
       \Bigr) [I,-A_k]^\Transp , \label{Qhat}
       \\[1ex] 
       \hat{R}_{k,j}& = (y_{k,j}-H\hat{m}_{k|T})(y_{k,j}-H\hat{m}_{k|T})^\Transp  + H \hP _{k|T} H^\Transp ,
\end{align}
\end{subequations}
and the ``hat'' variables are computed by the smoother 
with the previous iterand of $ \theta= [q, R, m_{1|0}, P_{1|0}]$.

Formulas \eqref{QhatRhat} can also be written  in the form
\begin{subequations}   \label{QhatRhatSR}
\begin{align}
   \hQ_k &= 
       (\hat{m}_{k+1|T}- A_k \hat{m}_{k|T}) (\hat{m}_{k+1|T}- A_k \hat{m}_{k|T})^\Transp \nonumber \\
         & \mbox{} \quad 
         +  (A_k G_k \hQ_k^{1/2}) (A_k G_k \hQ_k^{1/2})^\Transp  \nonumber \\
         & \mbox{} \quad 
      +   \bigl( (I-A_kG_k)(\hP _{k+1|T}^{1/2}+A_k \hP _{k|k}^{1/2}) \bigr)
       \nonumber \\
         & \mbox{} \quad  \quad
 \bigl( (I-A_kG_k)(\hP _{k+1|T}^{1/2}+A_k \hP _{k|k}^{1/2}) \bigr)^\Transp,
       \label{QhatSR}
       \\[1ex] 
         \hat{R}_{k,j}& = (y_{k,j}-H\hat{m}_{k|T})(y_{k,j}-H\hat{m}_{k|T})^\Transp  \nonumber
         \\ & \qquad \mbox{} + (H \hP _{k|T}^{1/2})(H \hP _{k|T}^{1/2})^\Transp .
\end{align}
\end{subequations}
Each term in  \eqref{QhatRhatSR} is a product of a matrix with its transpose.
Linear algebra software libraries include codes
to  compute  products of this form efficiently and with exact preservation of symmetry;
for example
 the multiplication operator \texttt{*} in \textsc{Matlab} is overloaded to do this.


\subsection{Dense output}


 The posterior estimate of the state at an inter-abscissa time $t_{k+\theta}=t_k+\theta\Delta_k$, with $0<\theta<1$,
  conditional on the measurements at times up to and including $t_k$,
  is  denoted
  \[
    x_{k+\theta | k} \sim \Normal(m_{k+\theta|k},P_{k+\theta|k}).
 \]
Its parameters can be obtained using the RTS smoother forward pass formulas
by omitting the measurement update stage 
 and applying a  dynamic update stage with the modified dynamic model
 \begin{subequations}  \label{interp1}
 \begin{align}
 m_{k+\theta|k} &=A_{k,\theta} m_{k|k}   \label{minterp1}  \\
 P_{k+\theta|k} &= A_{k,\theta} P_{k|k}A_{k,\theta}^\Transp + Q_{k,\theta} \label{Pinterp1}
 \end{align}
 \end{subequations}
where
$A_{k,\theta} = \exp(F\theta \Delta_k) $
and $ Q_{k,\theta}=q\bar{Q}_{k,\theta}$ with 
\[
\bar{Q}_{k,\theta}  = \int_0^{\theta\Delta_k} \exp(F(\theta\Delta_k-\tau))LL^\Transp  \exp(F^\Transp (\theta\Delta_k-\tau)) \ud \tau.
\]
That is, the formulas for the modified model matrices are obtained by using $\theta \Delta_k$
in place of   $\Delta_k$ in the formulas for the dynamic transition matrix and process noise covariance
given earlier.

The posterior estimate of the interpolatory state conditional on \emph{all} the measurements,
\[     x_{k+\theta|T} \sim \Normal( m_{k+\theta|T},P_{k+\theta|T}), \]
is obtained using  the backward pass formula
to go from $t_{k+1}$ to $t_{k+\theta}$ instead of to  $t_{k}$:
\begin{subequations}  \label{interp}
\begin{align}
G_{k,\theta} &= P_{k+\theta| k}A_{k,1-\theta}^\Transp  P_{k+1|k}^{-1}  \label{Ginterp}  \\
m_{k+\theta | T} &= m_{k+\theta|k}+G_{k,\theta}(m_{k+1|T}-m_{k+1|k})  \label{minterp}
\end{align}\end{subequations}

The functional form of the interpolant can be inferred from these formulas.
Substituting \eqref{interp1} and \eqref{Ginterp} into \eqref{minterp} gives
\begin{align*}
& m_{k+\theta   | T} = A_{k,\theta} m_{k|k}
\\ & \qquad \mbox{}+(A_{k,\theta} P_{k|k}A_{k,\theta}^\Transp + Q_{k,\theta})
A_{k,1-\theta}^\Transp  P_{k+1|k}^{-1}  (m_{k+1|T}-m_{k+1|k}) .
\end{align*}
Because the coefficients of $A_{k,\theta}$, $A_{k,1-\theta}$ and $Q_{k,\theta}$
are polynomials in $\Delta_k$,
 so is the interpolant $m_{k+\theta | T}$.
In particular,  its first component (the displacement) 
is a polynomial of degree $2d-1$.

\subsection{Initial  parameters}
Although EM has good theoretical convergence properties,
the convergence can be slow.
This slowness can be offset by making a reasonably good choice of initial parameter values.
In the \textsc{Matlab} implementation, 
the initial iterands for the state $m_{1|0}$ and 
the measurement noise variance $R$ are set by least-squares fitting 
a straight line through the first 10 abscissas.
The covariance $P_{1|0}$ is set to a tiny multiple of the identity matrix.
The driving noise intensity $q$ is then set by minimizing the negative log likelihood,
 a univariate minimization  whose cost function \eqref{eq:phi}
is computed using a Kalman filter.


\section{Tests}


Corradini et al. \cite{corradini93} compare ND algorithms using five test functions
that resemble experimental measurements of
different kinds of human movement.
They considered different measurement noise levels and sampling rates,
and found  no large differences in accuracy between the
five algorithms that they tested. 
They however identify two  algorithms,
which they label F1 and F2,
 as being the most accurate:
the smoothing heptic spline of~\cite{woltring86} 
(widely used  because its code is freely available)
and the fixed-lag Kalman smoother of~\cite{fioretti89} with three states.
These are also the only algorithms in their tests that are automatic,
except that the measurement noise variance needs to be specified.

Table~\ref{table:corradini} shows the errors of displacement, velocity, and acceleration
estimates reported in \cite{corradini93} 
for 94-point noisy displacement sequences generated from five test functions.
 The  error of the estimate of the derivative sequence
 is reported as  the percentage of RMS error relative to the true sequence's RMS value.
Also shown are the errors found with the proposed  algorithm
with $d=3$ states.
 The EM iterations were repeated until the norm of the change in the displacement estimate
was less than 0.1\% of the norm of the estimate;
no more than 3 EM iterations were needed in any of the tests.

The methods' errors are not precisely comparable,
because different random number generators were used to produce the 
measurement noise for the data sequences.
However, the results indicate that
the accuracies of  the proposed method
are roughly as good and in some cases clearly better than those
of the reference methods.

\begin{table}
\centering
\begin{tabular}{lrrrrrrrrr}
test &  method & displ. & vel. & accel. \\ \hline
T1 &  F1 &0.14 & 7.27 & 45.5 \\[-1ex]
    &   F2  & 0.13 & 6.10 & 36.8 \\[-1ex]
     &   new & 0.15 & 2.99 & 11.1 \\ 
T2  &  F1& 3.51 & 9.64 & 25.9 \\[-1ex]
    &  F2 & 3.55 & 9.94 & 26.7  \\[-1ex]
     &   new & 2.64 & 8.28 & 24.8 \\ 
T3 &  F1 & 3.02 & 9.13 & 26.1 \\[-1ex]
     & F2 & 3.04 & 9.40 & 26.9  \\[-1ex]
     &   new & 2.26 & 9.37 & 24.3 \\ 
T4 &  F1 & 2.32 & 10.40 & 30.7  \\[-1ex]
      & F2 & 2.39 & 10.83 & 34.3  \\[-1ex]
     &   new & 1.77 & 8.22 & 33.4 \\ 
T5 &  F1 & 1.95 & 10.34 & 39.5 \\[-1ex]
     &   F2 & 1.87 & 9.27 & 36.0 \\[-1ex]
     &   new & 1.12 & 6.42 & 20.6 \\ 
\end{tabular}
\caption{Relative RMS errors (in percentage) of estimates
using synthetic displacement data generated from five test functions. \label{table:corradini}}
\end{table}

\section{Conclusions}

The algorithm presented here is based on the integrated Wiener process,
which as argued in~\cite{fioretti94} is a principled and flexible signal model
for estimation of derivatives from noisy time series.
The proposed ND algorithm has  some advantages 
over that of \cite{fioretti94}:
it uses a numerically stable square-root smoother algorithm,
allows non-equally spaced and simultaneous data,
and its implementation is freely available.
Also, the  ML  parameters are  
computed using a reliable EM iteration,
which gives an automatic ND algorithm 
whose accuracy is as good or better than other  methods.

The assumption of additive Gaussian  noise 
may be inadequate for measurements with sporadic outliers.
This shortcoming could be addressed by replacing the RTS smoother by
 a Student-t smoother~\cite{piche12}. 


%


\bibliographystyle{plain}
\bibliography{numder}{}

\appendix
\section{Derivation of EM update formulas}
Substituting 
the  state space model's data log-likelihood    
\begin{align*}
  \log p(x_{1:T},y_{1:T}\cond \theta)
 &  = \log p(x_1\cond \theta)+\sum_{k=1}^{T-1}\log p(x_{k+1}\cond x_k ,\theta)
 \\ & \mbox{}
  + \sum_{k=1}^T \sum_{j=1}^{n_k}  \log p(y_{k,j}\cond x_k,\theta)
\end{align*}
into 
the EM objective function
\[ 
\mathcal{Q}(\theta,\hat{\theta}) = \int p(x_{1:T}\cond y_{1:T},\hat{\theta})\log p(x_{1:T},y_{1:T}\cond \theta)\ud x_{1:T}
\] 
(where $\hat{\theta}$ is the previous iteration's parameter value) gives
\begin{align*}
\mathcal{Q}&(\theta,\hat{\theta}) 
  =  \int p(x_1 \cond y_{1:T}, \hat{\theta}) \log p(x_1\cond \theta)  \ud x_1 
\\ & \quad \mbox{}+ \sum_{k=1}^{T-1}   \int  p(x_{k+1},x_{k} \cond y_{1:T}, \hat{\theta}) \log p(x_{k+1}\cond x_{k},\theta) \ud x_{k+1}   \\ & \quad \mbox{} + \sum_{k=1}^T  \sum_{j=1}^{n_k} \int p(x_k\cond y_{1:T},\hat{\theta}) \log p(y_{k,j}\cond x_k,\theta) \ud x_{k}
\ud x_{k}.
\end{align*}
This is a sum of expectations of log terms.
From \eqref{ssm_initial}, \eqref{ssmProcess}, \eqref{ssmMeasurement}, 
the log terms are
\begin{align*}
 \log p&(x_1\cond \theta)=-\Half \log \det (2\pi P_{1|0}) 
 \\ & \qquad \mbox{} -\Half (x_1-m_{1|0})^\Transp  P_{1|0}^{-1} (x_1-m_{1|0}),
 \\
  \log p&(x_{k+1}\cond x_{k},\theta) =-\Half \log \det (2\pi q\bar{Q}_k) 
     \\ & \qquad \mbox{}
      -\Half (x_{k+1}-A_kx_k)^\Transp  (q\bar{Q}_k)^{-1} (x_{k+1}-A_kx_k),
  \\
   \log p&(y_{k,j}\cond  x_k,\theta) =-\Half \log \det (2\pi R) 
   \\ & \qquad \mbox{}
        -\Half (y_{k,j}-Hx_k)^\Transp  R^{-1} (y_{k,j}-Hx_k).
\end{align*}
The distributions with respect to which the expectations 
are taken are
\begin{align*}
x_1 & \sim \Normal (\hat{m}_{1|0},\hP _{1|0}), \\
\begin{bmatrix} x_{k+1} \\ x_{k} \end{bmatrix}
& \sim \Normal\Bigl(   
\begin{bmatrix} \hat{m}_{k+1|T} \\ \hat{m}_{k|T} \end{bmatrix},
\begin{bmatrix} \hP _{k+1|T} & \hP _{k+1|T}\hG_k^\Transp  \\ \hG_k\hP _{k+1|T} & \hP _{k|T} \end{bmatrix}
\Bigr),
\\
x_k & \sim \Normal (\hat{m}_{k|T},\hP _{k|T}),
\end{align*}
where hats indicate values that are computed by the smoothing algorithm
applied to the model having parameters 
$\hat{\theta}= [\hat{q}, \hat{R}, \hat{m}_{1|0}, \hP _{1|0}]$.
Computing the expectations gives the formula for the EM objective function as
\begin{align*}
\mathcal{Q}(\theta,\hat{\theta}) &= -\Half \log \det (2\pi {P}_{1|0})
     - \Half \Tr\bigl(P_{1|0}^{-1} \hP _{1|T}\bigr) 
     \\ & \quad \mbox{}
      - \Half 
     (\hat{m}_{1|T}-m_{1|0})^\Transp  P_{1|0}^{-1}(\hat{m}_{1|T}-m_{1|0})
\\ & \quad \mbox{} -  \Half \sum\limits_{k=1}^{T-1} \Bigl( \log \det (2\pi Q_k) 
   + \Tr \bigl( Q_k^{-1}   \hQ_k \bigr) \Bigr)
\\ & \quad \mbox{} -  \Half \sum\limits_{k=1}^T  \sum\limits_{j=1}^{n_k}
         \Bigl( \log \det (2\pi R) + \Tr \bigl( R^{-1}   \hat{R}_{k,j} \bigr) \Bigr),
\end{align*}
where $\hQ_k$ and $\hat{R}_{k,j}$ are given by  \eqref{QhatRhat}.
Using standard matrix differential calculus formulas~\cite{magnus07}, 
the partial derivatives of the EM objective function  are 
\begin{align*}
\partial \mathcal{Q}(\theta,\hat{\theta})/\partial q &=
 \Half \sum\limits_{k=1}^{T-1} \Tr\Bigl( Q_k^{-1} \frac{\partial Q_k}{\partial q} \bigl(-I+\hQ_kQ_k^{-1} \bigr)\Bigr)
 \\
 &= \Half\Bigl( -\frac{(T-1)d}{q} + \frac{1}{q^2} \sum\limits_{k=1}^{T-1} \Tr(\hQ_k \bar{Q}_k^{-1}) \Bigr),
\\
\partial \mathcal{Q}(\theta,\hat{\theta})/\partial R 
  &=   \Half \sum\limits_{k=1}^T  \sum\limits_{j=1}^{n_k}
  R^{-1}  \bigl(-I+ \hat{R}_{k,j}R^{-1}  \bigr),
\\
\partial \mathcal{Q}(\theta,\hat{\theta})/\partial m_{1|0}^\Transp  &= 
    (\hat{m}_{1|T}-m_{1|0})^\Transp  P_{1|0}^{-1}  ,
\\
\partial \mathcal{Q}(\theta,\hat{\theta})/\partial P_{1|0} &= \Half P_{1|0}^{-1}
   \Bigl(-I +\bigl( \hP _{1|T} \\
   & \quad \mbox{} +(\hat{m}_{1|T}-m_{1|0})(\hat{m}_{1|T}-m_{1|0})^\Transp \bigr)\Bigr) P_{1|0}^{-1} .
\end{align*}
Setting these  to zero and solving gives the EM update formulas~(\ref{EMformula}--\ref{QhatRhat}).

The covariance matrix in \eqref{Qhat} can be written as
\begin{align*}
&\begin{bmatrix} \hP _{k+1|T} &  \hP _{k+1|T}\hG_k^\Transp  \\ \hG_k  \hP _{k+1|T} &  \hP _{k|T} \end{bmatrix} 
\\\qquad &=
\begin{bmatrix} I & 0 \\ \hG_k & I \end{bmatrix}
\begin{bmatrix} \hP _{k+1|T} & 0 \\ 0 & \hP _{k|T}-\hG_k\hP _{k+1|T}\hG_k^\Transp \end{bmatrix}
\begin{bmatrix} I & 0 \\ \hG_k & I \end{bmatrix}^\Transp.
\end{align*}
Substituting the identities
\[ \hat{P}_{k|T} = \hat{P}_{k|k}+\hG_k(\hat{P}_{k+1|T}-\hat{P}_{k+1|k})\hG_k^\Transp \]
and
\[ \hat{P}_{k+1|k} = A_k \hat{P}_{k|k}A_k^\Transp +\hQ_k ,
\]
and applying the Joseph formula,
the last element of the diagonal matrix can be rewritten as
\begin{align*}
\hP _{k|T}-\hG_k\hP _{k+1|T}\hG_k^\Transp &= 
\hP _{k|k}-\hG_k\hP _{k+1|k}\hG_k^\Transp \\
&=\hP _{k|k}-\hG_k(A_k \hP _{k|k}A_k^\Transp + \hQ_k)\hG_k^\Transp
\\
&= (I-\hG_kA_k)\hP _{k|k}(I-\hG_kA_k)^\Transp + \hG_k\hQ_k\hG_k^\Transp.
\end{align*}
Formula \eqref{QhatSR} is then obtained by
replacing the covariance matrices by their Cholesky factorisations.


\end{document}